\documentclass[vecphys]{svmult}
\pdfoutput=1

\usepackage{graphicx}
\usepackage{rotating}
\usepackage{color}
\usepackage{multicol}
\usepackage{sidecap}
\usepackage[bottom]{footmisc}
\graphicspath{{figures/}}
\DeclareGraphicsExtensions{.pdf,.png}

\makeindex

\usepackage{amsmath}
\usepackage{amsfonts}
\usepackage{listings}

\definecolor{somegrey}{rgb}{.95,.95,.99}
\lstnewenvironment{CSource}[1][]{
\lstset{#1}
\lstset{frame=tb}
\lstset{numbers=left, numberstyle=\tiny, stepnumber=1}
\lstset{backgroundcolor=\color{somegrey}}
\lstset{language=C}
}{
}

\begin{document}

\title*{Multi-core architectures: Complexities of performance prediction and the impact of cache topology}
\titlerunning{Performance prediction and cache topology on multi-core}
\author{Jan Treibig \and Georg Hager \and Gerhard Wellein}
\authorrunning{J. Treibig et al.}
\institute{
J. Treibig $\cdot$ G. Hager $\cdot$ G. Wellein
\at Regionales Rechenzentrum Erlangen, Friedrich-Alexander Universit\"at Erlangen-N\"urnberg, 
Martensstr. 1, D-91058 Erlangen, Germany\\
\email\texttt{\{jan.treibig,georg.hager,gerhard.wellein\}@rrze.uni-erlangen.de}
}

\maketitle

\abstract{The balance metric is a simple approach to estimate the
performance of bandwidth-limited loop kernels.
However, applying the method to in-cache situations and modern
multi-core architectures yields unsatisfactory results. This paper
analyzes the influence of cache hierarchy design on performance
predictions for bandwidth-limited loop kernels on current mainstream 
processors. We present a diagnostic model with improved predictive
power, correcting the limitations of the simple balance metric. The
importance of code execution overhead even in bandwidth-bound
situations is emphasized. Finally we analyze the impact of
synchronization overhead on multi-threaded performance with
a special emphasis on the influence of cache topology. } \newpage

\section{Introduction}
%
Many algorithms are limited by bandwidth, meaning that the memory subsystem
cannot provide the data as fast as the arithmetic core could process it. One
solution to this problem is to introduce multi-level memory hierarchies with
low-latency and high-bandwidth caches, which exploit spatial and (hopefully)
temporal locality in an application's data access pattern. In many scientific
algorithms the bandwidth bottleneck is still severe, however. A popular way to
estimate the performance in such situations is the memory bandwidth balance
metric~\cite{schoenauer}. This metric can estimate loop kernel performance very
well on vector systems and previous generations of cache-based processors. We
will show why the balance model fails on recent processors (Intel Nehalem) and
for in-cache situations. To overcome these limitations we introduce a
diagnostic performance model based on the real cache architectures and data
transfer paths. The application of the model is demonstrated on elementary data
transfer operations (load, store and copy operations) and benchmarked on three
x86-type test machines. In addition, as a prototype for many streaming
algorithms we use the STREAM triad $\vec{A}=\vec{B}+\alpha * \vec{C}$, which
matches the performance characteristics of many real algorithms~\cite{jalby}.
We show multi-threaded bandwidth measurements on shared caches, providing
valuable data on saturation effects.

Besides the limitations of shared outer-level caches and main memory
bandwidth, another important issue can influence multi-threaded performance:
synchronization overhead. We present measurements
investigating the influence of cache topology, threading implementations, different
OpenMP implementations and thread count on synchronization overhead. 

This paper is organized as follows. Section~\ref{sec:machines} gives
an overview on the microarchitectures and technical specifications of
the test machines. In Section~\ref{sec:bandwidth} we first present
the original balance model as introduced in~\cite{schoenauer} and
demonstrate its limitations, using a simple vector triad and a 
Jacobi relaxation solver. We then use a thorough analysis of cache
hierarchies to develop our diagnostic model and elaborate on
in-cache saturation effects that may harm multi-threaded performance.
In Section~\ref{sec:multithread} we finally pinpoint 
synchronization overheads on shared and separate caches.

%
\section{Experimental test bed}
\label{sec:machines}
\begin{figure}[b]
  \centering
    \includegraphics[width=0.31\linewidth]{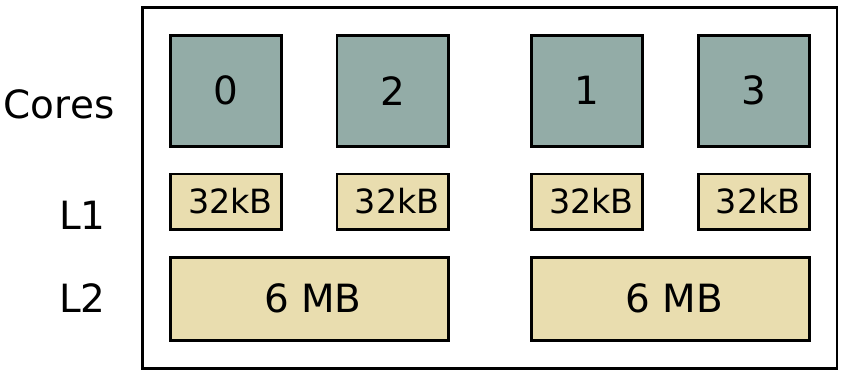}
    \includegraphics[width=0.31\linewidth]{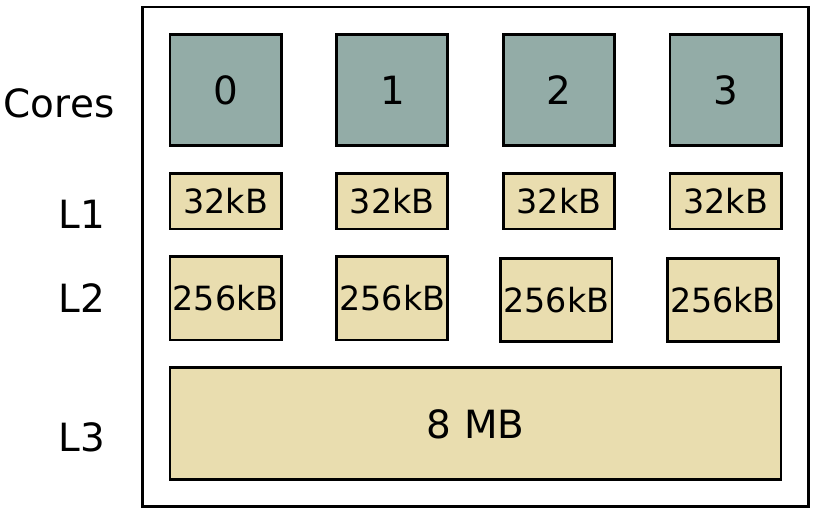}
    \caption{Cache group structure of the multi-core architectures in the
    test-bed for Core 2 (left) and Core i7 (right)}
    \label{fig:cache_arch}
\end{figure}

%
An overview of the machines used for benchmarking can be found in
Table~\ref{tab:arch}. As representatives of current x86 architectures we have
chosen Intel's ``Core 2 Quad'' and ``Core i7'' processors. The cache group
structure, i.e., which cores share caches of what size, is illustrated in
Figure~\ref{fig:cache_arch}. For detailed information about microarchitecture
and cache organization, see the Intel \cite{IntelOpt} Optimization Handbook.

Note that we will utilize two-socket variants of those systems in
Section~\ref{sec:multithread}, which are however very similar
on the one-socket level.
\begin{table}[tb]
    \caption{Test machine specifications. The cacheline size is 64 bytes 
	for all processors and cache levels.}
    \label{tab:arch}
    \centering
	\begin{tabular}{lccc}
	    \hline
	    &Core 2&Nehalem\\
	    &Intel Core2 Q9550&Intel i7 920\\
	    \hline
	    Execution Core&\\
	    \hline
	    Clock [GHz]&2.83&2.67\\
	    Throughput&4 ops&4 ops\\
	    Peak FP rate MultAdd&4 flops/cycle&4 flops/cycle\\
	    \hline
	    L1 Cache&32 kB     &32 kB   \\
	    Parallelism     &4 banks, dual ported &4 banks, dual ported\\
	    \hline
	    L2 Cache&2x6 MB (inclusive)&4x256 KB\\
	    \hline
	    L3 Cache (shared)&-&8 MB (inclusive)\\
	    \hline
	    Main Memory&DDR2-800&DDR3-1066\\
	    Channels&2&3\\
	    Memory clock [MHz]&800&1066\\
	    Bytes/ clock&16&24\\
	    Bandwidth [GB/s]&12.8&25.6\\
	    STREAM triad 1 thread [GB/s]&6.8&13.9\\
	    STREAM triad node [GB/s]&7.1&22.2\\
	    \hline
	\end{tabular}
\end{table}
%
\section{Bandwidth}
\label{sec:bandwidth}

\subsection{Memory Bandwidth Balance Model}
\label{sec:mem_balance}
%
The balance metric~\cite{schoenauer} sets into relation the number of data words a processor can
transfer from memory to the number of arithmetic operations it can execute.
This relation is also referred to as ``machine balance'', $B_M$. The
``algorithmic balance'' $B_A$ is the ratio between the number of words a given
algorithm needs per iteration to the number of arithmetic operations it
performs with this data. The expected efficiency (fraction of peak
performance) of the algorithm on a certain machine is then determined by the
relationship $\ell=\frac{B_M}{B_A}$.

\subsection{Limitations of the Memory Balance Model}
\label{sec:caches}

\subsubsection{In-cache Triad on the Core 2 architecture}
\label{sec:core2}
%
Even though the balance model was initially proposed for the memory domain, at a
first glance it seems to be worthwhile to also apply it to cache-bound
algorithms. As an example the STREAM triad on a Core 2 processor with 2.83 GHz is
considered. The L2 cache on the Core 2 has a 32-byte wide bus to the L1 cache
and runs with full clock speed. This results in a theoretical bandwidth of 
90.56~GBytes/s. The machine balance for the L2 cache is $B_M=\frac{11.32
\text{GWord/s}}{11.32 \text{GFlops/s}} \approx 1.0$\,Words/Flop. The algorithmic balance
for the STREAM triad is $B_A=\frac{3 \text{words}}{2 \text{flops}} \approx
1.5$\,Words/Flop. The expected efficiency is then $\frac{B_M}{B_A}=0.66$, resulting in
a prediction of 7.47~GFlops/s. 
However, measurements in the L2 domain yield only 1.99~GFlops/s.
Part of the discrepancy can be explained by the fact that Intel processors 
use a write allocate strategy for stores. Therefore a more
accurate prediction has to take into account the additional read for ownership (RFO)
for every store miss, resulting in an algorithmic balance of $B_A=\frac{4
\text{words}}{2 \text{flops}} \approx 2.0$\,Words/Flop. Now
$\frac{B_M}{B_A}=0.5$, resulting in a prediction of 5.66~GFlops/s, still
too large by a long shot. 

\begin{SCfigure}[1.47][tb]
    \includegraphics*[width=0.4\linewidth]{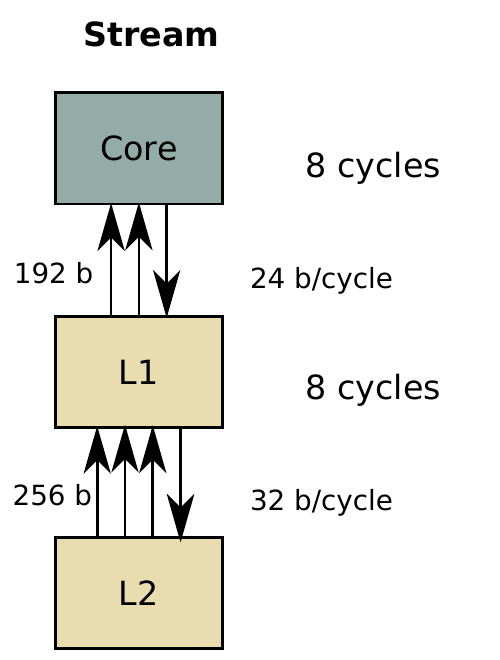}
    \caption{This figure shows cycles per cacheline
    update. A STREAM triad requires two load and one store instruction for one
    update. The Core~2 processor can execute either one 16-byte load and one
    16-byte store, one 16-byte load, or one 16-byte store per cycle. As
    there are more loads than stores, this leads to an effective bandwidth of 24
    bytes/cycle from L1 cache. For one cacheline update three cachelines have
    to be processed (192~bytes), therefore 8 cycles are needed. This analysis
    assumes that all other instructions can be executed in parallel. For the L2 cache,
    taking into account the RFO, four cachelines (256~bytes) have to be transferred, 
    leading to an additional 8~cycles. This is based on the L2 to L1 
    bandwidth of 32~bytes/cycle. Total runtime for one cacheline update 
    is therefore 16~cycles at minimum.}
    \label{fig:stream_core2_l2}
\end{SCfigure}

The reason why the balance metric fails for the in-cache case is that it
assumes that runtime is solely made up of the data transfers from the L2 cache.
Still in reality runtime is the sum of the cycles it takes to execute the
instructions with data located in the L1 cache and the cycles it takes to load
cachelines from L2 to L1 cache. These contributions cannot overlap, since
either the core's execution units or the cache controller can access the L1
cache at any given time. An analysis of the runtime contributions is shown in
Figure~\ref{fig:stream_core2_l2}. The analysis yields 16 cycles for a
cacheline update, while we measure 22.72~cycles. The difference is caused
by the data access latencies of the L2 cache. From the architectural
specifications, one would expect a latency of about 14 cycles,
but the Core 2 processor has a L2 to L1 hardware
prefetcher, which is able to limit the overhead to six cycles for four 
transferred cachelines.

\subsubsection{Jacobi smoother on the Nehalem architecture}
\label{sec:nehalem}
%
The stencil scheme for the Jacobi smoother uses an eight-point
update operation~\cite{datta08} (Listing~\ref{src:jacobi-source}).
\begin{CSource}[caption={Jacobi stencil code},label=src:jacobi-source] 
for(int i=1; i<n-1; i++) {                                                                                             
  for(int j=1; j<n-1; j++) {                                                                                           
    for(int k=1; k<n-1; k++) {                                                                                         
      tn[i][j][k] = frac * t[i][j][k] + frac * (
                             t[i-1][j][k] + t[i+1][j][k]                                                               
                           + t[i][j-1][k] + t[i][j+1][k]                                                               
                           + t[i][j][k-1] + t[i][j][k+1] );                                                            
    }                                                                                                                  
  }                                                                                                                    
}
\end{CSource}

This variant of the Jacobi smoother in three dimensions 
performs eight flops per update (six additions and two
multiplication)\@. The Nehalem node described in Section~\ref{sec:machines} was
used for the benchmarks. It is important to note that peak performance on this
architecture can only be achieved with an equal distribution between additions
and multiplications and full usage of packed SSE instructions. The peak main
memory bandwidth is 25.6~GBytes/s or 3.2 double precision GWords/s. This
results in a machine balance of $B_M=\frac{3.2 \text{GWords/s}}{10.64
\text{GFlops/s}} \approx 0.30$\,Words/Flop. This value is often considered as
an upper limit for memory-bound performance. At an algorithmic balance 
$B_A=\frac{8 \text{words}}{8 \text{flops}} \approx 1.0$\,Words/Flop a
performance of 3192~MFlops/s can be expected. In certain cases 
this can be within reasonable range
of real measurements. However, this is pure coincidence and caused by a
cancellation of two effects: The large memory bandwidth of the Nehalem
architecture as compared to L3 performance, and our ignorance towards the real
runtime contributions and data streams that have to be sustained from memory. As will be shown
in the following, care must be taken that the model is applied in a
sensible way.

The machine balance based on peak properties considers upper limits which
cannot be reached even in the theoretical case by the Jacobi algorithm. It is
necessary to adapt the machine balance to the algorithm under consideration. A
more realistic estimate is to consider the peak performance for the present
arithmetic instruction mix and the sustained single-threaded main memory
performance of the STREAM triad benchmark (as listed in Table
\ref{tab:arch}). This results in a new machine balance of $B_M=\frac{1.74
\text{GWord/s}}{6.65 \text{GFlops/s}} \approx 0.262$\,Words/Flop.  The initial
algorithmic balance is based on the properties of the Jacobi stencil update,
but can be significantly wrong on cache-based processors. Here the size of the
grid determines how many streams have to be loaded from main memory. As we
consider the balance model in the main memory domain, only the streams to
memory must be taken into account. 

\begin{SCfigure}[0.98][b]
    \centering
    \includegraphics[width=0.5\linewidth]{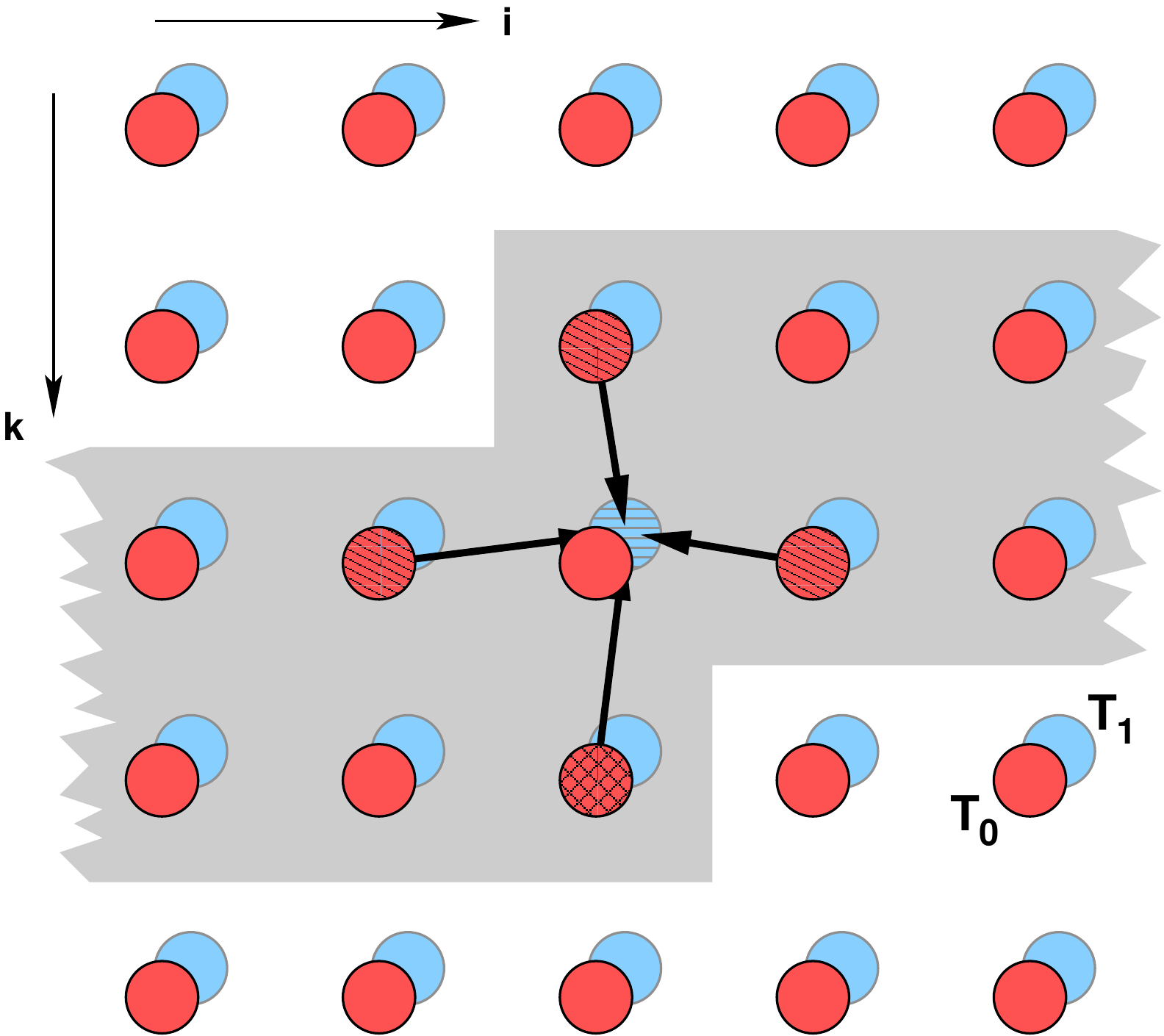}
    \caption{Schematic view of the Jacobi sweep in 2D using a five-point
    stencil. The shaded region marks the two grid rows (planes in 3D)
    that need to stay in cache in order to get three cache hits
    for the four loads.}
    \label{fig:jacobimem}
\end{SCfigure}

Depending on cache capacity, the data can be kept inside the caches between
multiple loads or is already evicted to memory. This is illustrated in
Figure~\ref{fig:jacobimem} using a 2D Jacobi algorithm. In three dimensions
the following cases can be distinguished: 
\begin{itemize}
\item If 
six grid rows (inner dimension) do not fit into the outer-level cache, this results in
six data streams from memory. This is the worst case scenario. 
\item If four planes
do not fit into the outer-level cache, four streams have to be loaded. 
\item With four complete planes fitting into cache only two streams have to come from
memory. 
\end{itemize}
The resulting machine balance and performance prediction based on the balance
metric is illustrated in Table~\ref{tab:stream} in the second and third
columns. In the last column measured performance is shown. It can be seen that
for six and four streams the prediction is accurate while for two streams the
performance is overestimated. This indicates that the simple balance model,
while accurate for situations with high pressure on the memory subsystem, fails
when many loads come from the outer-level cache. The first two cases are
unusual cases as only in rare cases four planes do not fit in the large 8 MB
L3 cache. Note that the Jacobi kernel was implemented
in assembly language without using non-temporal store instructions.

\begin{table}[tb]
    \centering
    \begin{tabular}{ccccc}
        streams  (with RFO)&~~$B_A$~~&~~$\ell$~~&~~predicted~~&~~measured~~\\
        \hline
        6 (7)&0.875&0.299&1988&1760 (88 \%)\\
        4 (5)&0.625&0.419&2786&2524 (90 \%)\\
        2 (3)&0.375&0.699&4648&3024 (65 \%)\\
    \end{tabular}
    \caption{Machine balance $B_A$ in Words/Flop and resulting prediction based on
    machine balance compared to the measured performance in
    MFlops/s for the three cases described in the text. Note that for all cases an additional stream 
    for the RFO is added.}
    \label{tab:stream}
\end{table}

A more realistic analysis of the Jacobi algorithm will be performed in
Section~\ref{sec:model_jacobi}, revealing the exact reasons for the balance
metric failing in certain situations. The main assumption of the balance metric
is that the contribution of in-cache data transfers and the execution of the
instructions can be neglected against the time required to transfer the data
from memory. The STREAM triad results (see Table~\ref{tab:arch}) are used as
the memory bandwidth component in the machine balance. The triad has a certain
relation between runtime spent on-chip and runtime used to transfer data from
main memory. In case of the Jacobi algorithm this relation is different and
also depends on the ratio between the number of data streams toward main memory
and cache.

Using the STREAM triad as the absolute limit for memory performance
is only justified for kernels which have similar on-chip contributions
to overall runtime, or on systems with very bandwidth-starved memory access.
The latter is not the case on the Nehalem architecture as will be shown
in Section \ref{sec:model_jacobi}. 



\subsection{Diagnostic performance model for bandwidth-limited loop kernels}
\label{sec:model}
%
The conclusions drawn from the simple kernel benchmarks in
Section~\ref{sec:caches} will now enable us to develop a diagnostic model for
loop kernels. This model proposes an iterative approach to analytically
predict the performance of bandwidth-limited algorithms in all memory hierarchy
levels. The basic building block of a streaming algorithm is its computational
kernel in the inner loop body. Since all loads and stores come from and go to
L1 cache, the kernel's execution time on a cacheline basis is governed by the
maximum number of L1 load and store accesses per cycle and the capability of
the pipelined, superscalar core to execute instructions. All lower levels of
the memory hierarchy are reduced to their bandwidth properties, with data paths
and transfer volumes based on the real cache architecture. The minimum
transfer size between memory levels is one cacheline.

Based on the transfer volumes and bandwidths, the total execution time per
cacheline is obtained by adding all contributions from data transfers between
caches and kernel execution times in L1\@. To lowest order, we assume that
there is no access latency (i.e., all latencies can be effectively hidden by
software pipelining and prefetching) and that the different components of
overall execution time do not overlap. 

It must be stressed that a correct application of this model requires an
intimate knowledge of cache architecture and data paths. This information is
available from processor manufacturers \cite{IntelOpt}, but sometimes the level
of detail is insufficient for fixing all parameters, and relevant information
must be inferred from measurements.

%
\subsubsection{Theoretical Analysis}
\label{sec:theory}
%
In this section we substantiate the model outlined above by providing
the necessary architectural details for current Intel
processors. Using simple kernel loops, we derive performance
predictions which will be compared to actual measurements in the
following section. All results are given in CPU cycles per cacheline;
if $n$ streams are processed in the kernel, the number of cycles
denotes the time required to process one cacheline per stream.

As mentioned earlier, basic data operations in L1 cache are limited by cache
bandwidth, which is determined by the load and store instructions that can
execute per cycle. The Intel cores can retire one 128-bit load and one 128-bit
store in every cycle. L1 bandwidth is thus limited to 16 bytes per cycle if
only loads (or stores) are used, and reaches its peak of 32 bytes per cycle
only for a copy operation.

For load-only and store-only kernels, there is only one data stream,
i.e., exactly one cacheline is processed at any time. With copy
and stream triad kernels, this number increases to two and three,
respectively. Together with the execution limits described above it
is possible to predict the number of cycles needed to execute the instructions
necessary to process one cacheline per stream
(see the ``L1'' columns in Table~\ref{tab:cache_model})\@.
\begin{table}[tb]
    \caption{Theoretical prediction of execution times for eight loop
	iterations (one cacheline per stream) on Core 2 (A) and Core i7 (B) processors}
    \label{tab:cache_model}
    \centering\renewcommand{\arraystretch}{1.15}\addtolength{\tabcolsep}{1mm}
\begin{tabular}{r|cc|cc|c|cc}
          &\multicolumn{2}{c|}{L1}&\multicolumn{2}{c|}{L2}&\multicolumn{1}{c|}{L3}&\multicolumn{2}{c}{Memory}\\
          &A  &B              &A   &B            &B                 &A     &B       \\
    \hline
    Load  &4  &4              &6   &6            &8                 &20    &15      \\
    Store &4  &4              &8   &8            &12                &36    &26      \\
    Copy  &4  &4              &10  &10           &16                &52    &36      \\
    Triad &8  &8              &16  &16           &24                &72    &51      \\
\end{tabular}
\end{table}

L2 cache bandwidth is influenced by three factors: (i) the finite bus width
between L1 and L2 cache for refills and evictions, (ii) the fact that
\emph{either} ALU access \emph{or} cache refill can occur at any one time, and
(iii) the L2 cache access latency. Both architectures have a 256-bit bus
connection between L1 and L2 cache and use a write back and write allocate
strategy for stores. In case of an L1 store miss, the cacheline is first moved
from L2 to L1 before it can be updated (write allocate)\@. Together with its
later eviction to L2, this results in an effective bandwidth requirement of 128
byte per cacheline write miss update. 

On Intel processors, a load miss incurs only a single cacheline transfer
from L2 to L1, because the cache hierarchy is inclusive. The Core i7 L2 cache is
not strictly inclusive, but for the benchmarks covered here (no cacheline
sharing and no reuse) an inclusive behavior was assumed due to the lack of
detailed documentation about the L2 cache.

The overall execution time of the loop kernel on one cacheline per stream is
the sum of (i) the time needed to transfer the cacheline(s) between L2 and L1
and (ii) the runtime of the loop kernel in L1 cache.
Table~\ref{tab:transfersum} shows the different contributions for pure load,
pure store, copy and triad operations on Intel processors. Looking at,
e.g., the copy operation, the model predicts that only 6 cycles out of
10 can be used to transfer data from L2 to L1 cache. The remaining 4 cycles are
spent with the execution of the loop kernel in L1. This explains the well-known
performance breakdown for streaming kernels when data does not fit into L1 any
more, although the nominal L1 and L2 bandwidths are identical.
All results are included in the ``L2'' columns of Table~\ref{tab:cache_model}\@.

Not much is known about the L3 cache architecture on Intel Core i7. It can be
assumed that the bus width between the caches is 256 bits, which was confirmed
by our measurements. Our model assumes a strictly inclusive cache hierarchy for
the Intel designs, in which L3 cache is ``just another level''.
Under these assumptions, the model can predict the required number of cycles in
the same way as for the L2 case above. The ``L3'' column in
Table~\ref{tab:cache_model} show the results.

\begin{figure}[tb]\centering
    \includegraphics[width=0.9\linewidth]{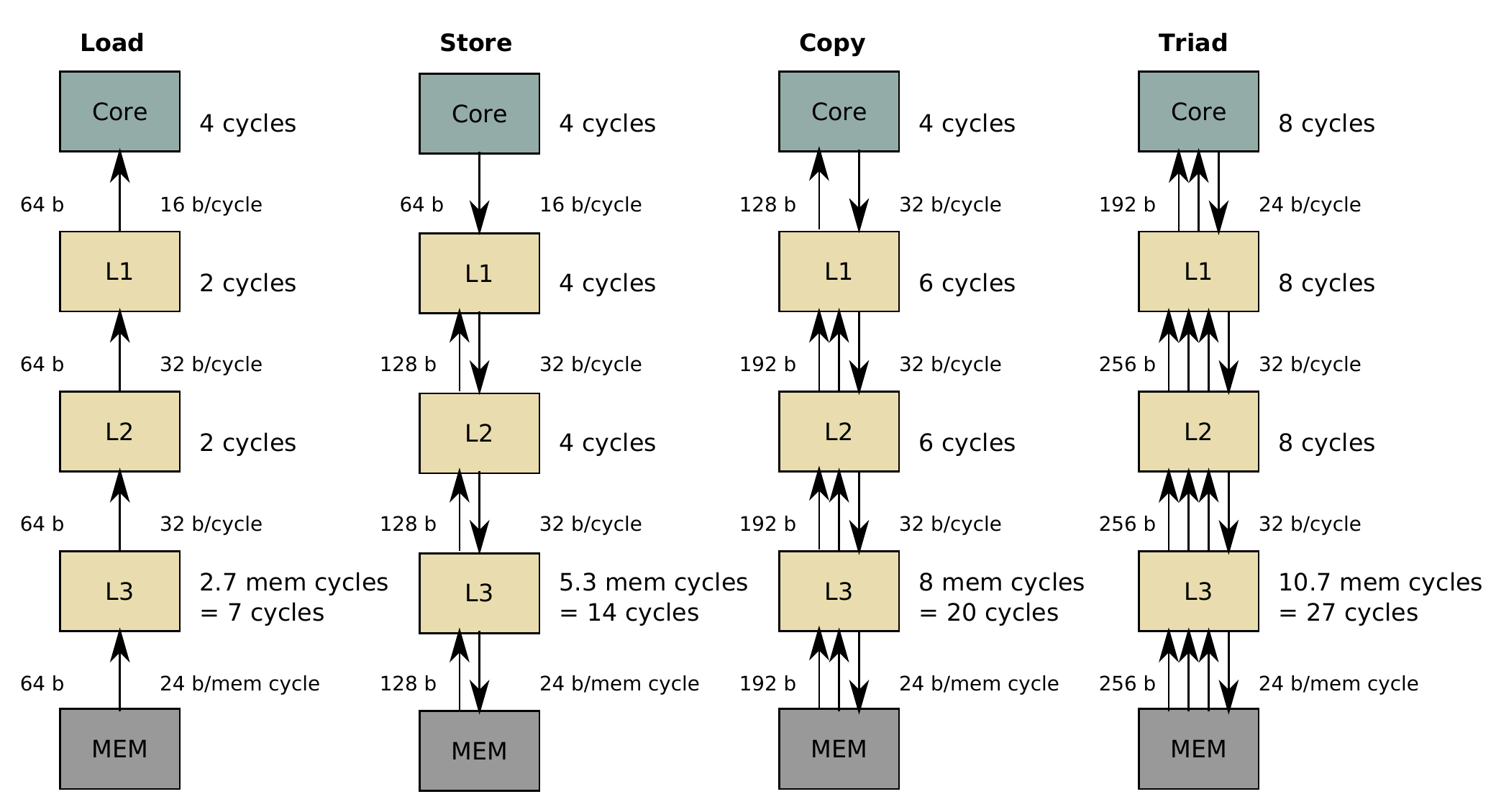}
    \caption{Main memory performance model for Intel Core i7.
    There are separate buses connecting the different cache levels.}
    \label{fig:mem_data_intel_nehalem}
\end{figure}

If data resides in main memory, we again assume a strictly hierarchical
(inclusive) data load on Intel processors.
The cycles for main memory transfers are computed using the effective memory
clock and bus width and are converted into CPU cycles. For consistency reasons, 
non-temporal (``streaming'') stores were not used for the main memory regime.
Data transfer volumes and rates, and predicted cycles for a cacheline update
are illustrated in Figure~\ref{fig:mem_data_intel_nehalem}.
 They are also included in the
``Memory'' columns of Table~\ref{tab:cache_model}\@.
%
\subsubsection{Measurements}
\label{sec:results}
%
Measured cycles for a cacheline update, the ratio of predicted versus measured
cycles, and the real and effective bandwidths are listed in
Table~\ref{tab:cache_results} for Load, Store, Copy and Triad
benchmarks. Here, ``effective bandwidth'' means the
bandwidth available to the application, whereas ``real bandwidth'' refers to
the actual data transfer taking place. For every layer in the hierarchy the
working set size was chosen to fit into the appropriate level, but not into
higher ones. The measurements confirm the predictions of the model well in the
L1 regime.

Also the L2 results confirm the predictions. One exception is the store
performance of the Intel Core i7, which is significantly better than the
prediction. This indicates that the model does not describe the store behavior
correctly. At the moment we have no additional information about the L2
behavior on Core i7 to solve this problem. The overhead for accessing the L2
cache with a streaming data access pattern scales with the number of involved
cachelines, as can be derived from a comparison of the measured cacheline
update cycles in Table~\ref{tab:cache_results} and the predictions in
Table~\ref{tab:cache_model}\@. The highest cost occurs on the Core 2 with 2
cycles per cacheline for the triad, followed by Shanghai with 1.5 cycles per
cacheline. Core i7 has a very low L2 access overhead of 0.5 cycles per cache
line. Still, all Core i7 results must be interpreted with caution until the L2
behavior can be predicted correctly by a revised model. Both architectures are
good at hiding cache latencies for streaming patterns.

On Core i7 the behavior with regard to the L3 cache is similar to the L2
results: The store result is better than the prediction, which influences all
other test cases involving a store. It is obvious that the Core i7 applies an
unknown optimization for write allocate operations.
\begin{table}[p]
    \caption{Benchmark results and comparisons to the predictions of the diagnostic model
    for Load, Store, Copy and Triad kernels}
    \label{tab:cache_results}
    \centering\footnotesize
    \begin{sideways}\renewcommand{\arraystretch}{1.5}
	\begin{tabular}{r|cccc|cccc|cccc|cccc}
	               &\multicolumn{4}{c|}{L1}       &\multicolumn{4}{c|}{L2}        &\multicolumn{4}{c}{L3}           &\multicolumn{4}{c}{Memory}        \\
	               &Load  &Store  &Copy   &Triad &Load   &Store  &Copy   &Triad &Load   &Store   &Copy   &Triad  &Load     &Store    &Copy       &Triad  \\
	    \hline                                                                                                                                                       
	    \multicolumn{1}{l|}{Core 2 [\%]}&96.0  &93.8   &92.7   &99.5   &83.1   &94.1   &74.9   &70.4   &&&&                              &67.6     &49,9      &58.7      &66.6    \\
	    CL update  &4.17  &4.26   &4.31   &8.04   &7.21   &8.49   &13.34  &22.72  &&&&                              &29.60    &72.04     &88.61     &108.15  \\
	    GB/s       &43.5  &42.5   &84.1   &67.7   &25.1   &42.7   &40.7   &31.9   &&&&                              &6.1      &5.0       &6.1       &6.7     \\
	    eff. GB/s  &-     &-      &-      &-      &-      &21.3   &27.2   &23.9   &&&&                              &-        &2.5       &4.1       &5.0     \\
	    \hline                                                                                                                                                      
	    \multicolumn{1}{l|}{Nehalem [\%]}&97.1  &95.3    &94.1   &96.0   &83.5   &120.9  &91.4   &91.7   &95.3   &121.4    &103.9   &96.3    &106.8    &142.2     &123      &119.4  \\
	    CL update   &4.12  &4.20    &4.26   &8.34   &7.18   &6.61   &10.94  &17.45  &8.39   &9.88    &15.4   &24.91   &14.02    &18.27     &29.25    &42.72  \\
	    GB/s        &41.3  &40.5    &79.8   &61.2   &23.7   &51.5   &46.7   &39.0   &20.3   &34.4    &33.2   &27.3    &12.1     &18.6      &17.4     &15.9   \\
	    eff. GB/s   &-     &-       &-      &-      &-      &25.7   &31.1   &29.3   &-      &17.2    &22.1   &20.5    &-        &9.3       &11.6     &11.9   \\
	    \hline                                                                                                                                                    
	\end{tabular}%
      \end{sideways}
\end{table}

As for main memory access, one must distinguish between the classic frontside
bus concept as used with all Core 2 designs, and the newer architectures with
on-chip memory controller. The former has much larger overhead, which is why
Core 2 shows mediocre efficiencies of around 60\,\%\@. The Core i7 shows
results better than the theoretical prediction on all memory levels except L1.
This might be caused either by a potential overlap between the different
contributions (which is ignored by our model), or by deficiencies in the model
caused by insufficient knowledge about the details of data paths inside the
cache hierarchy.
%

\subsubsection{Application to the Jacobi smoother on Nehalem}
\label{sec:model_jacobi}
\begin{SCfigure}[1.47][tb]
    \includegraphics*[width=0.4\linewidth]{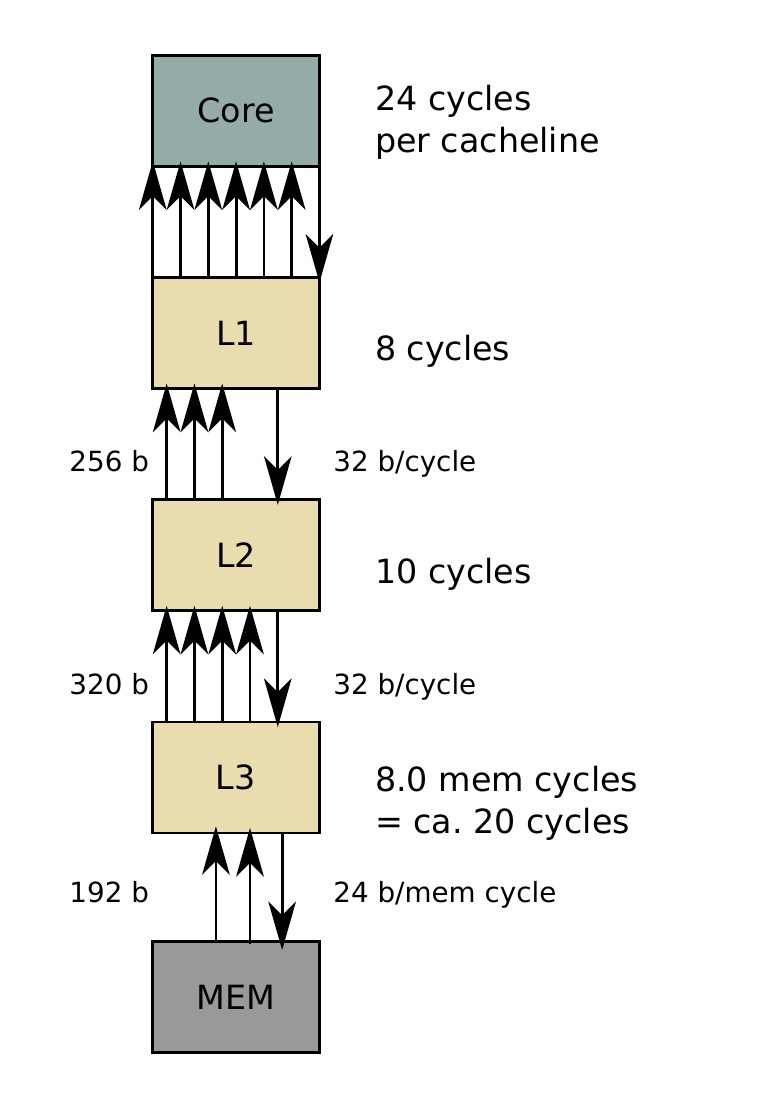}
    \caption{An instruction analysis shows that for 3D Jacobi 24 cycles are
    needed to update one cacheline. Here it is assumed that four planes fit
    into the L3 cache and seven lines fit into the L1 cache. This results in
    the cacheline transfers shown. Every arrow is one 64 byte
    cacheline transfer. Bus width between cache levels is 256 bit or a data
    transfer rate of 32 bytes/cycle. For the data rate to memory, the memory clock
    and memory bus width is taken into account. This results into 8 cycles to
    transfer the necessary cachelines from L2 to L1 cache, 10 cycles to
    transfer the cachelines from L3 and L2 cache and 20 cycles to load the
    cachelines from main memory into the L3 cache. Total runtime for one cache
    line update according to the model are 62 cycles.}
    \label{fig:jacobiModel}
\end{SCfigure}

%
An analysis of the Jacobi algorithm is shown in Figure~\ref{fig:jacobiModel}.
The prediction based on this analysis for the case with three data streams to
memory is 2745~MFlops/s, while we measure 3024~MFlops/s. It must be noted that
two important details on the Nehalem processor are not documented: First,
measurements show that runtime is overestimated if an RFO transfer from L2 to
L1 is assumed for each store miss.  This issue was already taken into account
in the model analysis in Figure~\ref{fig:jacobiModel}, and indicates that the
L1/L2 hierarchy is not accurately described by a simple inclusive structure.
Second, the possibility to overlap data transfers in different hierarchy levels
is neglected in our model. The measured performance indicates that our
prediction overestimates the time required by a cacheline update by six cycles.
Since these predictions were based on bandwidth capabilities we must conclude
that indeed the cache hierarchy allows an overlap between data transfers
and execution of the instructions from L1 cache.


Finally, the comparison between the model predictions and the measured
performance show that on the Nehalem architecture it is a good approximation
for stream-oriented algorithms to neglect data access latencies.

\subsection{Bandwidth saturation with shared caches}
\label{sec:saturation}
%
Shared caches are the ``glue'' holding together today's multi-core architectures.
While shared caches allow very fast data exchange and synchronization between 
cores, one possible drawback is that all cores share the bandwidth.

In order to measure the bandwidth saturation behavior of the shared caches, a
special version of a Load and Copy benchmark was implemented which only loads
and/or copies one 4-byte item per cacheline. This minimizes the influence of
instruction execution in the core and runtime is reduced to the pure data
transfers between cache hierarchies. The Core~2 and Core~i7 systems described
in Section~\ref{sec:machines} were used for all measurements.

The runtime for running the single-threaded loop kernel in the L1 domain was
chosen as the baseline. Although both CPUs are quite similar on this level, an
important difference is that the L1 cache of Core 2 has a lower latency than on
Nehalem (see Table~\ref{tab:transfersum})\@. From an architectural point of
view it should always be possible to execute the loop body in one cycle, a
target which both systems miss by a considerable amount. This is, however, not
unusual for very small loop bodies with high instruction throughput. 

Nehalem has less overhead for transfers between L1 and L2, saving 1 cycle per
cacheline against Core~2 for pure loads, and almost 3 cycles for copies. This
indicates a more efficient prefetching mechanism from L2 to L1 than on
Core~2\@. L2 performance on Nehalem is in certain cases better than predicted
based on the bandwidth capabilities (at least 4 cycles for Load and 10 cycles
for Copy as indicated in Figure~\ref{fig:mem_data_intel_nehalem}), suggesting
that the data paths on this level are not fully understood yet. 

Running two threads on separate L2 domains, both processors scale as expected
because there is no shared bandwidth. On a shared L2 (only possible with
Core~2), the Load benchmark on Core~2 loses about 1.2 cycles on average per
cacheline versus the non-shared case (note that measurements indicate that the
overhead roughly scales with the number of transferred cachelines)\@. For the
shared Copy benchmark the loss is nearly 4 cycles. Core 2 has a theoretical
transfer limit of 32~bytes/cycle between L2 and L1, which results in a
bandwidth of 90.56~GB/s on our test machine. Translating the cycle measurements
to effective bandwidths, it turns out that one thread cannot saturate the L2
bus: It achieves 36.6~GB/s for Load and 44.2~GB/s (including RFO) for Copy.
With two threads on the shared cache, these numbers increase to 57.4~GB/s and
63.3~GB/s, respectively. Hence, although there is some headroom for providing
additional bandwidth from L2 to the second core, scalability is far from
perfect. It would thus not make sense to have a Core~2 design with more than
two cores on a shared L2\@.

An additional third cache level can decouple loads and stores from L3-L2
refills and enlarge the time slots available to transfer data from the L3 cache
for each core. However, a large L3 cache usually has larger latencies as well.
Nehalem compensates this by very effective prefetchers which achieve L3
bandwidths similar to (if not better than) the L2 cache on Core~2. Our
measurements show that Nehalem's L3 can fully scale up to two cores for Copy
and to three cores for Load, with considerable headroom for an additional core.
It reaches its peak L3 bandwidth at 85~GB/s (Load) and 63~GB/s (Copy)\@.
Latency effects cannot be detected, suggesting that Nehalem overlaps transfers
between L3 and L2 with the execution of loads and stores from L1\@. 

\begin{table}[tb]\centering
    \caption{Shared cache load and copy benchmark [cycles per cacheline update]} 
    \label{tab:transfersum}
    \addtolength{\tabcolsep}{1mm}
\begin{tabular}{r|cc|cc}
    &\multicolumn{2}{c|}{Core 2}&\multicolumn{2}{c}{Core i7}\\
                            &Load   &Copy    &Load  &Copy\\
    \hline
    L1 1 thread             &1.45   &2.25    &2.21  &2.24\\
    \hline
    L2 1 thread             &4.95   &12.29   &3.85  &9.47\\
    L2 2 threads non-shared &4.97   &13.01   &3.82  &9.51\\
    L2 2 threads shared     &6.31   &17.16   &-     &-   \\
    \hline
    L3 1 thread             &-      &-       &5.26  &14.98\\
    L3 2 threads shared     &-      &-       &5.33  &16.86\\
    L3 3 threads shared     &-      &-       &6.05  &24.48\\
    L3 4 threads shared     &-      &-       &8.01  &33.35\\
\end{tabular}
\end{table}

\section{Synchronization effects in multi-core environments}
\label{sec:multithread}
An elementary component in many multi-threaded algorithms is
synchronization between different threads. Commonly, a barrier is used
to ensure synchronized execution. The available x86 multi-core
processors have no hardware support for synchronization primitives,
hence software solutions are required. Depending on the granularity
of synchronization, this can incur significant overhead, as will be
shown in the following.

We compare three different options for barrier synchronization:
\begin{itemize}
    \item The OpenMP barrier (Intel icc~11.0 and GNU gcc~4.3.3 compiler)\@.
      The barrier benchmark contained in the EPCC OpenMP Microbenchmarks V2.0
      \cite{bull} was used for this.
    \item The pthread (NPTL~2.9) barrier.
    \item A spin-waiting loop, implemented following the guidelines
      from the Intel Optimization Handbook~\cite{IntelOpt} and using
      dedicated cachelines for all synchronization variables.
\end{itemize}
The original EPCC code was endowed with cycle-accurate timing. An
equivalent benchmark was implemented for the pthread and spin-waiting
loop variants.

Note that we mainly present measured results here, to be used as
guidelines when estimating synchronization overheads on the
architectures and software environments under consideration. Without
intimate knowledge about the underlying OpenMP and pthread
implementations, it is next to impossible to find the exact reasons
for the deviations observed.

\subsection{Two threads on a shared cache}

The first case we consider is the interaction of two threads sharing a 
cache. Table~\ref{tab:barrier} shows the complete results.

On a Core~2-based quad-core processor (see Section~\ref{sec:machines}) the spin
waiting loop needs 231~cycles to synchronize the two threads. The Intel OpenMP
implementation is significantly slower (399~cycles)\@. In contrast, the
pthread barrier and the OpenMP implementation provided by gcc are around 60
times slower, both taking over 20000~cycles per synchronization. (Thus gcc can
be expected to provide especially low performance on code with relatively short
parallel loops or regions because of the implicit barriers imposed by OpenMP.)
A look in the source code of both the NPTL pthread and gcc OpenMP
implementation reveals that both rely on the \verb+futex+ system
call~\cite{futex}.
\begin{table}[tb]
    \centering
    \begin{tabular}{l|cc}
        &Intel Q9550 (sh. L2)&Intel i7 920 (sh. L3)\\
        \hline
        pthreads\_barrier\_wait&23739  &6511\\
        omp barrier (icc~11.0) &399    &469\\
        omp barrier (gcc~4.3.3)&22603  &7333\\
        spin waiting loop      &231    &270\\
    \end{tabular}
    \caption{Runtime of different thread barrier primitives (in CPU cycles)
      for two threads in a shared cache}
    \label{tab:barrier}
\end{table}
On the Core~i7 the results for the spin waiting loop and the Intel OpenMP
barrier are slightly worse than on Core~2\@. On the other hand, the pthread
and gcc OpenMP barriers improve to around 7000~cycles, reducing the difference
to a factor of 14\@. The reason for this discrepancy could be that pthread and
gcc barrier overhead are dominated by the inefficient \verb.futex. mechanism,
while the more efficient Intel OpenMP and spin waiting loops sense the slightly
larger cache latencies on Nehalem as compared to Core~2\@.

In summary, Intel's OpenMP barrier implementation provides reasonable 
synchronization performance on a shared cache, and is only outperformed
by an optimized spin wait. Pthread and gcc OpenMP barriers appear to
use a very inefficient underlying mechanism.

\subsection{Cache and node topology}

Placing two threads on cores with separate caches and/or separate sockets
measures the influence of node topology. The results are shown in
Table~\ref{tab:barrier_topology}. Note that these tests were conducted on
dual-socket machines.
From now on we omit results for the gcc barrier because it uses the same
underlying mechanism as pthread, and the performance results are very similar.
Note that the pthread results cannot be compared to the results in the previous
section. Because of the complex influences of the kernel and the pthread library
results on different machines (even with the same processor architecture) show 
a large variance in the results.

For the Core~2 architecture, good results are achieved by the Intel OpenMP
implementation with 576~cycles on a shared cache, 760~cycles on one socket, and
1269~cycles on different sockets. The pthread barrier, apart from being much
less efficient even on the shared cache, loses a factor of four if the threads
run in separate caches. While the spin waiting loop reaches the best overall
result on a shared cache and on separate caches inside a socket, a striking
performance loss occurs on different sockets.
\begin{table}[tb]
    \centering
    \begin{tabular}{l|ccc}
        Intel Xeon E5420&~~shared L2 cache&same socket&different socket\\
        \hline
        pthreads\_barrier\_wait &5863  &27032 &27647\\
        omp barrier (icc 11.0)~~  &576   &760   &1269\\
        spin waiting loop       &259   &485   &11602\\
        \hline
        Intel Nehalem&SMT threads&shared L3 cache&~different socket\\
        \hline
        pthreads\_barrier\_wait &23352 &4796  &49237\\
        omp barrier (icc 11.0)  &2761  &479   &1206\\
        spin waiting loop       &17388 &267   &787\\
        \hline
    \end{tabular}
    \caption{Topology influence on thread barrier primitives for 
	two threads (in CPU cycles) }
    \label{tab:barrier_topology}
\end{table}

On the Nehalem architecture the behavior of the pthread barrier and Intel
OpenMP is comparable to Core~2\@. Note that the spin waiting loop is relatively
efficient on separate sockets with 787~cycles, probably because cachelines can
be exchanged directly across the QuickPath link without explicit eviction to
memory. The influence of simultaneous multi-threading (SMT) is of special
interest, so we also considered synchronization between two threads on the same
physical core but different logical cores. There are already considerable
penalties for both the Intel OpenMP and pthread barriers (about a factor of 6),
but the spin waiting loop loses a factor of 65 because of severe resource
contention between the SMT threads. This effect is well-known from former SMT
implementations (called ``Hyper-Threading'' on Pentium~4 processors)\@.

In summary, again the Intel OpenMP barrier yields satisfactory average
performance across all topology variants on Core~2\@. Our spin wait loop, while
outperforming Intel on one socket, obviously misses an important optimization
aspect when synchronizing different sockets. On the other hand, the Nehalem
architecture seems to be well suited for spin-waiting, except when running
threads on the same physical core, which is, to varying degree, never a good
idea, regardless of the synchronization method. If SMT must be used, Intel's
OpenMP barrier is clearly the best solution. An possible solution to the low
SMT and dual socket performance of the spin waiting loop is to replace the 
spin loop with a mechanism which senses the synchronization objects of other cores
only every few hundred cycles halting the core in the meantime. This would increase 
the response time for very short synchronization periods but prevent the problems 
in the SMT case and for multiple sockets leading to overall more balanced results.

\subsection{Barrier cost for many threads}

In Table~\ref{tab:barrier_count} we show the scaling of barrier cost when using
all threads on one versus two sockets. On one socket the spin waiting loop
achieves best results on both architectures, as could be expected from the
two-threads results above. For Core~2 the overhead roughly doubles when
including the second socket, but the impact is much larger on Nehalem.
However, if threads must be synchronized across the whole node, Intel OpenMP
and spin waiting are roughly on par. The former is still dominating, however,
if SMT threads are used.

Overall, pthread and gcc OpenMP synchronization are not suited for fine-grained
parallelization. For the ease of use and overall balanced results the Intel
OpenMP barrier is the preferred solution. The spin waiting loop reaches best
results for shared caches but is outperformed on different sockets (Core~2) and
SMT threads (Nehalem)\@. 

As a rule of thumb, synchronizing all threads in a two-socket node costs of the
order of one microsecond. Barriers on SMT threads should be avoided by all
means.
\begin{table}[tb]
    \centering
    \begin{tabular}{l|ccc}
        Intel Xeon E5420&~~4 threads&~~8 threads&\\
        \hline
        pthreads\_barrier\_wait&31436&60664&\\
        omp barrier (icc 11.0)~~&1290&2040&\\
        spin waiting loop&1084&2761\\
        \hline
        Intel Nehalem&4 threads&8 threads&~~16 threads SMT\\
        \hline
        pthreads\_barrier\_wait&10355&58577&89635\\
        omp barrier (icc 11.0)&794&2373&5431\\
        spin waiting loop&448&1915&20033\\
        \hline
    \end{tabular}
    \caption{Overhead for thread barrier primitives when scaling from 
	one to two sockets (in CPU cycles)}
    \label{tab:barrier_count}
\end{table}

\section{Conclusion and Outlook}
\label{sec:conclusion}
%

Using single-threaded, stream-based benchmarks and a Jacobi solver, we have demonstrated
why performance modeling by a simple balance metric fails for current
multi-core architectures (Intel Core~2, Nehalem) and in-cache situations. Based
on these results we have introduced a diagnostic performance model, which led
to a deeper understanding of runtime contributions in all memory hierarchy
levels and finally to more accurate predictions. However, lacking some
important details about the covered microarchitectures, up to now not all
observed effects can yet be explained by the model.

Using load and copy microbenchmarks, the outer level cache bandwidth scaling
behavior was analyzed to answer the question whether shared caches may impose
bandwidth bottlenecks. While not all effects on the Nehalem could be explained
in-depth it was clearly shown what the differences are between Core~2 and
Nehalem and how the shared L3 cache scales in a multi-threaded context for
load- and copy-type streaming patterns.

Finally as a major performance-limiting issue for multi-threaded codes
synchronization overhead was analyzed. Different
synchronization primitives were compared and the influence of cache/node
topology and thread count on synchronization overhead was measured. The results
may help to decide how to best utilize the complex, hierarchical CPU and node
topologies in production environments.

Future work will include a thorough analysis of multi-threaded interleaving effects
for shared caches and memory. We will substantiate our findings with performance
counter measurements and develop tools which allow even end users to gain a deeper
understanding of their applications' bandwidth behavior.

\section*{Acknowledgments}

We thank Darren Kerbyson (LANL), Herbert Cornelius (Intel Germany),
Michael Meier (RRZE), and Matthias
M\"uller (ZIH) for fruitful discussions. This work was financially
supported by the KONWIHR-II project ``Omi4papps''.

%

\bibliographystyle{spphys}

\end{document}